\documentclass[twocolumn,superscriptaddress]{revtex4-2} 
\setcitestyle{super}
\usepackage[T1]{fontenc}
\usepackage{times}
\usepackage{graphicx}

\usepackage{amsfonts, amsmath,amssymb, bm, graphicx}
\usepackage{siunitx}

\usepackage{amssymb}
\usepackage[english]{babel}
\usepackage{lipsum}

\newcommand{\figref}[2]{\hyperref[#1]{\ref{#1}(#2)}}
\usepackage{physics}
\usepackage{lipsum}
\usepackage{changes}

\usepackage[colorlinks=true,allcolors=blue]{hyperref}
\usepackage{footmisc}

\usepackage{upgreek}

\usepackage{soul}

\makeatletter
\let\ORIbbl@fixname\bbl@fixname
\def\bbl@fixname#1{%
  \@ifundefined{languagealias@\expandafter\string#1}
    {\ORIbbl@fixname#1}
    {\edef\languagename{\@nameuse{languagealias@#1}}}%
}
\newcommand{\definelanguagealias}[2]{%
  \@namedef{languagealias@#1}{#2}%
}
\makeatother

\definelanguagealias{en}{english}

\begin{document}

{
\makeatletter
\def\frontmatter@thefootnote{%
 \altaffilletter@sw{\@fnsymbol}{\@fnsymbol}{\csname c@\@mpfn\endcsname}%
}%

\makeatother
\title{Numerical reverse engineering of general spin-wave dispersions:\\ Bridge between numerics and analytics using a dynamic-matrix approach}

\author{L. K\"orber}\email{l.koerber@hzdr.de}
\affiliation{Helmholtz-Zentrum Dresden - Rossendorf, Institut f\"ur Ionenstrahlphysik und Materialforschung, D-01328 Dresden, Germany}
\affiliation{Fakultät Physik, Technische Universit\"at Dresden, D-01062 Dresden, Germany}

\author{A. Kákay}
\affiliation{Helmholtz-Zentrum Dresden - Rossendorf, Institut f\"ur Ionenstrahlphysik und Materialforschung, D-01328 Dresden, Germany}

\date{\today}

\begin{abstract}
Modern problems in magnetization dynamics require more and more the numerical determination of the spin-wave spectra and -dispersion in magnetic systems where analytic theories are not yet available. Micromagnetic simulations can be used to compute the spatial mode profiles and oscillation frequencies of spin-waves in magnetic system with almost arbitrary geometry and different magnetic interactions. Although numerical approaches are very versatile, they often do not give the same insight and physical understanding as analytical theories. For example, it is not always possible to decide whether a certain feature (such as dispersion asymmetry, for example) is governed by one magnetic interaction or the other. Moreover, since numerical approaches typically yield the normal modes of the system, it is not always feasible to disentangle hybridized modes. In this manuscript, we build a bridge between numerics and analytics by presenting a methodology to calculate the individual contributions to general spin-wave dispersions in a fully numerical manner. We discuss the general form of any spin-wave dispersion in terms of the effective (stiffness) fields produced by the modes. Based on a special type of micromagnetic simulations, the numerical dynamic-matrix approach, we show how to calculate each stiffness field in the respective dispersion law, separately for each magnetic interaction. In particular, it becomes possible to disentangle contributions of different magnetic interactions to the dispersion asymmetry in systems where non-reciprocity is present. At the same time, dipolar-hybridized modes can be easily disentangled. Since this methodology is independent of the geometry or the involved magnetic interactions at hand, we believe it is attractive for experimental and theoretical studies of magnetic systems where there are no analytics available yet, but also to aid the development of new analytical theories.
\end{abstract}

\maketitle

\section{Introduction}

Within the research field of magnonics, which studies the dynamics of spin waves (magnons)\cite{blochZurTheorieFerromagnetismus1930} in magnetically-ordered media, one of the key interests is to determine the dispersion and spatial mode profiles of spin waves with wavelengths in the nanometer to micrometer regime. Since the second half of the 1960s, a lot of works have been devoted to develop analytical theories, which allow the dipole-exchange spectrum in ferromagnetic films,\cite{sparksEffectExchangeMagnetostatic1970,sparksFerromagneticResonanceThin1970a,wolframEffectExchangeMagnetic1970,wolframMagnetoexchangeBranchesSpinWave1971a,camleyTheoryLightScattering1981,ariasExtrinsicContributionsFerromagnetic} with arguably on of the most successful and, because of its convenience, one of the most widely-used theories being the model developed by Kalinikos and Slavin based on magnetostatic Green's functions.\cite{kalinikosTheoryDipoleexchangeSpin1986} From thin films, the literature of spin-wave propagation has since has been expanded to more complicated geometries, such as magnetic bilayers,\cite{gallardoReconfigurableSpinWaveNonreciprocity2019} waveguides with rectangular cross section,\cite{guslienkoEffectiveDipolarBoundary2002a,kostylevDipoleexchangePropagatingSpinwave2007} systems with Dzyaloshinskii-Moriya interactions (DMI),\cite{cortes-ortunoInfluenceDzyaloshinskiiMoriya2013,moonSpinwavePropagationPresence2013} magnetic nanotubes\cite{otaloraCurvatureInducedAsymmetricSpinWave2016,otaloraAsymmetricSpinwaveDispersion2017,otaloraFrequencyLinewidthDecay2018} or nanowires,\cite{ariasTheorySpinExcitations2001a,gaidideiLocalizationMagnonModes2018,rychlySpinWaveModes2019} and many more.\cite{gladiiFrequencyNonreciprocitySurface2016,gaidideiMagnetizationNarrowRibbons2017}

With the recent advancements in high-performance computing, full numerical approaches like time-domain micromagnetic simulations or dynamic-matrix approaches are used more and more to calculate spin-wave spectra in complex systems where analytical (or semi-analytical) theories are not yet available. Such systems include, in particular, more complex geometries as well as systems with sufficiently inhomogeneous equilibrium magnetization.

Typical time-domain simulations rely on the numerical time-integration of the nonlinear Landau-Lifshitz-Gilbert equation of motion and, due to their free availability, are the most commonly used numerical method in magnetization dynamics.\cite{donahueOOMMFUserGuide1999,kakaySpeedupFEMMicromagnetic2010, vansteenkisteDesignVerificationMuMax32014} The spin-wave spectra can be obtained from such simulations by means of Fourier analysis. As an alternative method, numeric dynamic-matrix approaches directly yield the mode profiles and frequencies by solving the linearized equation of motion with a numerical eigensolver.\cite{grimsditchMagneticNormalModes2004,daquinoComputationMagnetizationNormal2012} Over the last two decades, dynamic-matrix approaches have been developed for various classes of problems and geometries, including confined magnetic structures,\cite{grimsditchMagneticNormalModes2004,daquinoComputationMagnetizationNormal2012,brucknerLargeScaleFiniteElement2019} rectangular waveguides, extended slabs and multilayers \cite{henryPropagatingSpinwaveNormal2016} and, recently, waveguides with arbitrary cross section.\cite{korberFiniteelementDynamicmatrixApproach2021} 

Both methods, time-domain and dynamic-matrix approaches, can be extremely powerful, and, due to their flexibility, applicable to almost arbitrary geometries, inhomogeneous equilibrium states and involved magnetic interactions. However, as a major disadvantage, both numerical approaches output the oscillation frequencies $\omega$ as plane numerical values, providing no insight into the functional structure of the dispersion $\omega = \omega(\bm{k})$. For example, without an (even approximate) analytical form of the dispersion, it can be hard to disentangle asymmetric contributions to the spin-wave dispersion at hand or even to uncover whether a certain effect is dominated by one magnetic interaction (such as exchange) or the other (such as dipolar interaction). Sometimes, qualitative statements can be found by switching off certain interactions or fields in the simulations. However, often times, the physical interpretation of a calculated spectrum solely based on the numerical values of the oscillation frequencies -- or on the \textit{shape} of the dispersion curves -- is up to the experience of the scientist. 

As a more subtle disadvantage, full numerical approaches do not allow to easily disentangle hybridized modes. In the field of magnonics, common termology is to denote different propagating spin-wave modes at a given wave vector $\bm{k}$ by their spatial profile. For example, in magnetic films, the spin waves with a certain in-plane wave-vector $\bm{k}_\parallel$ are often divided into a magnetostatic surface wave (MSSW) and several perpendicular-standing spin waves (PSSWs).\cite{damonMagnetostaticModesFerromagnet1961,stancilSpinWavesTheory2009a} Similarly, in thin rectangular waveguides, the modes with a given wave number $k$ along the waveguide can by divided by their standing-wave character along the width of the waveguide.\cite{roussigneExperimentalTheoreticalStudy2001,guslienkoEffectiveDipolarBoundary2002a} However, due to dynamic dipolar fields, these modes are often not actual eigenmodes (or normal modes) of the magnetic systems but are rather mixed by what is commonly referred to as dipole-dipole hybridization. Typically, hybridization removes certain points of accidental degeneration and produces avoided level crossings [see Fig.~\figref{fig:intro}{a}]. For example, in thick magnetic slabs, it is well known that dynamic dipolar fields may lead to mixing/hybridization of the MSSW and different PSSWs,\cite{qinExchangetorqueinducedExcitationPerpendicular2018,klinglerSpinTorqueExcitationPerpendicular2018,tacchiStronglyHybridizedDipoleexchange2019} schematically shown in Fig.~\figref{fig:intro}{a}.

Typical micromagnetic simulations, which have not been designed in a problem-specific way, always output the already dipole-dipole-hybridized normal modes of the system. When categorizing different modes in a numerically obtained spin-wave dispersion, knowledge about which modes are hybridized can be paramount.

\begin{figure}[h!]
    \centering
    \includegraphics{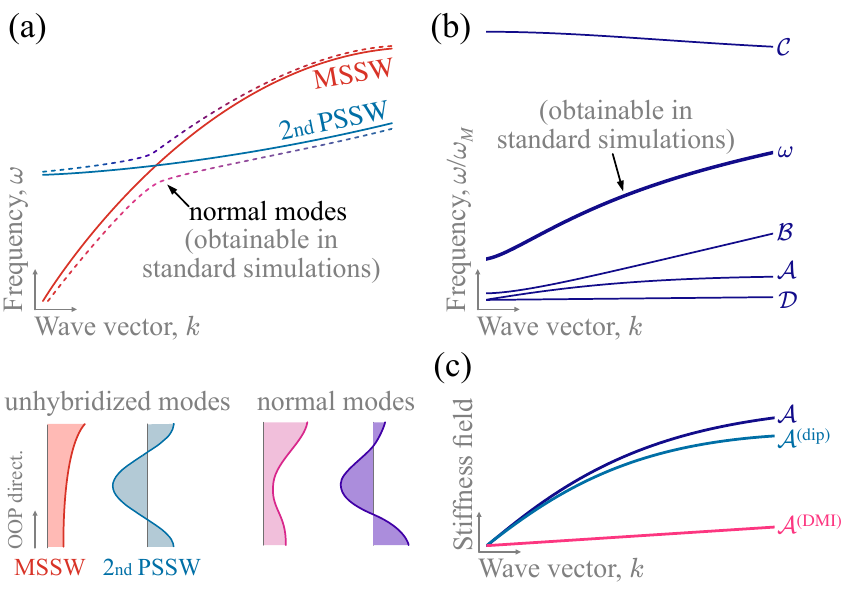}
    \caption{(a) Schematics of an avoided level crossing of the MSSW and the second PSSW in a magnetic slab which are hybridized due to dynamic dipolar fields. The mode profiles below represent a dynamical component of the mode along the thickness or out-of-plane (OOP) direction of the slab. (b) Exemplary spin-wave dispersion relation $\omega(k)$ comprised of the different stiffness fields $\mathcal{A}$, $\mathcal{B}$, $\mathcal{C}$ and $\mathcal{D}$. The magnetochiral stiffness field $\mathcal{A}$ is resolved in terms of the different magnetic interactions in (c). }
    \label{fig:intro}
\end{figure}

In the present paper, we present a fully numerical methodology to solve both of the aforementioned issues at the same time using dynamic-matrix approaches. As a theoretical groundwork, based on effective spin-wave tensors, we discuss that any spin-wave dispersion can be written in the form 
\begin{equation}\label{eq:general-dispersion-intro}
    \frac{\omega}{\omega_M} = \mathcal{A} + \sqrt{\mathcal{B}\mathcal{C}-\mathcal{D}^2}
\end{equation}
with $\omega_M =\mu_0\gamma M_\mathrm{s}$, where $\gamma$ is the gyromagnetic ratio, $M_\mathrm{s}$ is the saturation magnetizion of the material at hand, and $\mathcal{A}$, $\mathcal{B}$, $\mathcal{C}$ and $\mathcal{D}$ are the (unitless) dynamic stiffness fields. Specifically, $\mathcal{A}$ is the magnetochiral stiffness field which is responsible for any dispersion asymmetry (upon reversal of the wave vector). We will show that each of these terms can be obtained separately in a fully numerical fashion [see Fig.~\figref{fig:intro}{b}], both in terms of hybridized and non-hybridized modes. Moreover, contributions of different magnetic interactions can be easily disentangled. For example
\begin{equation}
    \mathcal{A} = \mathcal{A}^{(\mathrm{dip})} + \mathcal{A}^{(\mathrm{DMI})} + \hdots
\end{equation}
with $\mathcal{A}^{(\mathrm{dip})}$ and $\mathcal{A}^{(\mathrm{DMI})}$ being contributions of the dipolar interaction and the Dzyaloshinskii-Moriya interaction (DMI) to the dispersion asymmetry [Fig.~\figref{fig:intro}{c}]. Note, that the fully numerical approach does neither require any \textit{a priori} knowledge about the geometry of the magnetic element, nor about the involved magnetic interactions. We believe that our method is of particular interested to theoretically study spin-wave propagation in systems where there is no analytical model available. It may help in interpreting both experiments and micromagnetic simulations in complex geometries, but also aid the development of new analytical theories.

For this purpose, in Sec.~\ref{sec:dynmat}, we quickly review the basics of linear magnetization dynamics, and introduce the numerical dynamic-matrix approach. Moreover, we briefly derive the general form Eq.~\eqref{eq:general-dispersion-intro} based on effective spin-wave tensors. Following, in Sec.~\ref{sec:methods} we present how to recover unhybridized spin-wave dispersions as well we the individual terms in the respective dispersion using fully numerically obtained mode profiles. Finally, in Sec.~\ref{sec:methods}, we demonstrate this methodology by recovering the unhybridized dispersion in a standard magnonic waveguide, as well as by obtaining the individual stiffness fields in two further examples. The second example consists of a system, which combines dipolar- with bulk-DMI-induced dispersion asymmetry.

\section{Theoretical model}\label{sec:dynmat}

\subsection{Linearized equation of motion and dynamic matrix}
The problem of determining the spin-wave dispersion and spatial mode profiles for wavelengths far greater than the lattice parameter in a given magnetic specimen comes down to solving the linearized Landau-Lifshitz equation of motion\cite{stancilSpinWavesTheory2009a}
\begin{equation}
\label{eq:dynamicEq}
\dv{}{t} \qty(\delta \bm{m})=-\omega_M \big[\delta\bm{m}\times \bm{h}_0 + \bm{m}_0\times \delta\bm{h}\big]
\end{equation}
which describes the linear dynamics of the unitless small-amplitude variation $\delta\bm{m}(\bm{r},t)$ of the magnetization around some stable equilibrium direction $\bm{m}_0(\bm{r})$. The full magnetization of the sample is given as $\bm{M}(\bm{r},t) \approx M_\mathrm{s}(\bm{m}_0(\bm{r}) +\delta\bm{m}(\bm{r},t))  $. In the absence of surface anisotropy, the linearized equation must be equipped with the exchange boundary condition 
\begin{equation}\label{eq:boundarycond}
    \bm{n}\cdot \nabla\bm{m} = 0 
\end{equation}
and the orthogonality condition 
\begin{equation}\label{eq:orthogality}
    \delta\bm{m} \cdot \bm{m}_0 = 0
\end{equation}
the latter is a consequence of the micromagnetic constraint\cite{brownjr.Micromagnetics1963} $\abs{\bm{M}}=M_\mathrm{s}$. In Eq.~\eqref{eq:dynamicEq}, the vectors $\bm{h}_0$ and $\delta\bm{h}$ are the unitless static and dynamic effective fields, respectively. Both fields can be expressed as 
\begin{equation}\label{eq:eff}
    \bm{h}_0 =  \bm{h}_\mathrm{ext}- \vu{N}\cdot\bm{m}_0\qquad \delta \bm{h} = - \vu{N}\cdot\delta\bm{m}
\end{equation}
with $\bm{h}_\mathrm{ext}=\bm{H}_\mathrm{ext}/M_\mathrm{ext}$ being the unitless static external magnetic field and $\vu{N}$ being the magnetic tensor which is a certain self-adjoint integro-differential operator describing the magnetic self interactions, such as uniaxial magnetocrystalline anisotropy, isotropic exchange, dipolar interaction or Dzyaloshinskii-Moriya interaction (DMI),
\begin{equation}
    \vu{N} = \vu{N}^{\mathrm{(uni)}} + \vu{N}^{\mathrm{(exc)}} + \vu{N}^{\mathrm{(dip)}} + \vu{N}^{\mathrm{(DMI)}} + ...
\end{equation}
For expressions, see for example Refs.~\citenum{korberFiniteelementDynamicmatrixApproach2021} and \citenum{verbaHamiltonianFormalismNonlinear2019}. The contributions to the magnetic tensor relevant for this paper are summarized in Appx.~\ref{appx:mattensor}. Note, that the affine linear relationship Eq.~\eqref{eq:eff} between magnetization and effective field is not possible for example for cubic magnetocrystalline anisotropy, which can, however, be considered using a linearization. By expressing the effective fields using the magnetic tensor $\vu{N}$ and expanding the dynamic magnetization into linear waves $\propto \exp(i\omega_\nu t)$ with angular frequency $\omega_\nu$ and some mode index $\nu$, the linearized equation can be brought to the form\cite{naletovIdentificationSelectionRules2011, verbaCollectiveSpinwaveExcitations2012,tarasshevchenkonationaluniversityofkyiv64volodymyrskastr.kyiv01601ukraineSpinWavesArrays2013} 
\begin{equation}\label{eq:eigenllg}
    \frac{\omega_\nu}{\omega_M}\bm{m}_\nu = i[\bm{m}_0 \times \vu{\Omega}]\bm{m}_\nu
\end{equation}
with the $\bm{m}_\nu \equiv \bm{m}_\nu(\bm{r})$ being the (complex-valued) unitless spatial mode profiles. The operator $\vu{\Omega}$ is given as
\begin{equation}
   \vu{\Omega} =  h_0 \vu{I} + \vu{N}.
\end{equation}
with $h_0 = \bm{h}_0 \cdot \bm{m}_0 $ and $\vu{I}$ being the identity operator. For mode profiles which satisfy the exchange and orthogonality conditions in Eqs.~\eqref{eq:boundarycond} and \eqref{eq:orthogality}, the linearized equation Eq.~\eqref{eq:eigenllg} takes the form of a standard Hermitian eigenvalue problem, which, in general, can be solved numerically. This is exactly the idea behind the so-called dynamic-matrix approach. In contrast to standard micromagnetic simulations, which rely on a time integration of the initial (non-linearized) Landau-Lifshitz-Gilbert equation of motion, the dynamic-matrix approach directly yields the spin-wave frequencies and mode profiles of the magnetic eigen modes (or normal modes) without any additional post-processing. 

For numerical applications, it proves useful to encode the two aforementioned constraints already into the mathematical form of the eigenvalue equation itself. For example, in the finite-element method, the exchange boundary condition is already satisfied by a proper choice of basis functions. The orthogonality condition $\bm{m}_\nu \perp \bm{m}_0$ can be included by rotating the eigenvalue problem from the lab system $\{\bm{e}_x, \bm{e}_y, \bm{e}_z\}$ into the subspace spanned by $\{\bm{e}_1, \bm{e}_2\}$ locally orthogonal to the equilibrium  direction $\bm{m}_0(\bm{r})$.\cite{daquinoComputationMagnetizationNormal2012} Then the eigenvalue problem becomes 
\begin{equation}
    \frac{\omega_\nu}{\omega_M} \Tilde{\bm{m}}_\nu = \vu{D}\Tilde{\bm{m}}_\nu
\end{equation}
with the mode profiles in the local coordinate system $\Tilde{\bm{m}}_\nu = \vu{R}\bm{m}_\nu$, the dynamic matrix $\vu{D}$ including the aforementioned rotation operator $\vu{R}$ (see Appx.~\ref{appx:dynmat}). After this transformation, the dynamic matrix is only a $2\times 2$ matrix and explicitly only acts on the two $\textit{dynamical}$ components of the mode profiles ${\bm{m}}\cdot\bm{e}_{1,2}$, which are, again, locally orthogonal to the equilibrium $\bm{m}_0$. In case the magnetic specimen is discretized using $n$ mesh nodes, the dynamic matrix is $2n\times2n$ dimensional, whereas the mode profiles in the local basis have $2n$ components. It follows, that a maximum of $2n$ eigenvectors can be obtained.

To calculate the dispersion of propagating spin waves, which are characterized by their wave vector $\bm{k}$ (or wave number $k$ for propagation in only w.l.o.g. the $z$ direction), one arrives at the so-called propagating-wave dynamic-matrix approach by making the replacements
\begin{align}
        \bm{m}_\nu \rightarrow \bm{\eta}_{\nu k}& = \bm{m}_\nu e^{-ikz}\\
        \vu{D}  \rightarrow \vu{D}_k &= e^{-ikz} \vu{D} e^{ikz}
\end{align}
The task of a numerical dynamic-matrix approach is now to discretize the dynamic matrix $\vu{D}_{(k)}$ and diagonalize it using a suitable numerical solver. Over the past decades, a number of implementations of this method can be found which have already been successfully applied to a large variety of problems, both for standing\cite{grimsditchMagneticNormalModes2004,daquinoComputationMagnetizationNormal2012,brucknerLargeScaleFiniteElement2019} and propagating \cite{henryPropagatingSpinwaveNormal2016,korberFiniteelementDynamicmatrixApproach2021} waves. The power of these approaches is that they yield the spin-wave dispersion and mode profiles directly for a large class of different geometries and interactions, since almost all information about the spin-wave propagation is contained in the magnetic tensor $\vu{N}_{(k)}$. In the present paper, we shall apply our method to a propagating-wave dynamic-matrix approach which we have developed recently to obtain the spin-wave dispersion and mode profiles in infinite waveguides of arbitrarily-shaped finite cross section.\cite{korberFiniteelementDynamicmatrixApproach2021} Note, however, that it can be applied to any dynamic-matrix approach (\textit{e.g.} by dropping the index $k$ in some of the equations below).

\subsection{General form of spin-wave dispersion}\label{sec:genera-dispersion}

In this section, we shall derive the general form of a spin-wave dispersion, as already sketched in Eq.~\eqref{eq:general-dispersion-intro}. Later, we will show how to numerically obtain each individual term. We shall proceed without an \textit{a priori} knowledge of the underlying magnetic material, geometry or involved interactions. That is, we assume some arbitrary (but stable) equilibrium magnetization $\bm{m}_0(\bm{r})$ and some (of course, Hermitian) magnetic tensor $\vu{N}$. Let us assume that the spatial mode profiles, \textit{i.e.} the eigenvectors of Eq.~\eqref{eq:eigenllg}, are also known as $\bm{m}_\nu(\bm{r})$. Here, the index $\nu$ may just be a simple index to count the eigenmodes of the system, but my also be a collection of indices. In the case of a quasi-infinite waveguide, one may take for example $\nu = (k, \alpha,\beta)$ with $k$ denoting the wave number along the waveguide and $\alpha$ and $\beta$ denoting two mode indices identifying the lateral dependence of the mode profiles. To find a general spin-wave dispersion, it proves useful to define the \textit{effective spin-wave tensors} $\vu{N}_\nu$ via
\begin{equation}
\begin{split}
   \vu{N}\sum\limits_\nu c_\nu(t)\, \bm{m}_\nu (\bm{r})= \sum\limits_\nu c_\nu(t)\, \vu{N}_\nu \bm{m}_\nu (\bm{r})
\end{split}
\end{equation}
as the \textit{eigenvalues} of the magnetic tensor with respect to the mode profiles. The formalism of effective spin-wave tensors has been introduced by Nazarov \textit{et al.}\cite{nazarovGeneralSpinWave2002} for plane waves and is used for example in the theory of nonlinear spin-wave dynamics.\cite{krivosikHamiltonianFormulationNonlinear2010,verbaHamiltonianFormalismNonlinear2019} However,  we do not restrict the mode profiles to be plane waves, necessarily. These tensors represent the effective magnetic field generated by each individual mode profile. Moreover, $\vu{N}_\nu$ only contain (in general complex) numbers and are not integro-differential operators anymore. This allow us to write the dynamic matrix 
for a given mode $\nu$ as
\begin{equation}
    {\vu{D}}_\nu = \mqty(-i N^{(21)}_\nu & - i\qty(N^{(22)}_\nu+h_0) \\ i\qty(N^{(11)}_\nu+h_0) & i N^{(12)}_\nu)
\end{equation}
with $N_{ij}$ ($i,j=1,2$) being the spin-wave tensor elements in the local basis $\{\bm{e}_1,\bm{e}_2\}$. Using the fact, that all $\vu{N}_\nu$ have to be Hermitian as well, \textit{i.e.} $N^{(21)} =(N^{(21)})^*$, we can obtain the general spin-wave dispersion as the eigenvalues of the dynamic matrix,
\begin{equation}\label{eq:general-dispersion}
\begin{split}
        \frac{\omega_\nu}{\omega_M} & = \Im N^{(21)}_\nu\\ & \quad + \sqrt{\qty(N^{(11)}_\nu+h_0)\qty(N^{(22)}_\nu+h_0) - \qty( \Re N^{(21)}_\nu)^2} \\ 
        & \equiv  \mathcal{A}_\nu + \sqrt{\mathcal{B}_\nu\mathcal{C}_\nu-\mathcal{D}_\nu^2}.
\end{split}
\end{equation}
This form is identical to the one \textit{e.g.} obtained by Cortés-Ortunño and Landeros \cite{cortes-ortunoInfluenceDzyaloshinskiiMoriya2013,landerosTwoMagnonScattering2008} for uniform films with bulk DMI. Moreover, for $N_{21}=0$, it reduces to the famous Kittel formula.\cite{kittelTheoryFerromagneticResonance1948} 

Even without imposing any specific magnetic equilibrium, material, or involved interactions, we can make further statements about the general form of Eq.~\eqref{eq:general-dispersion} for the case of propagating waves, where \textit{e.g.} $\vu{N}_\nu = \vu{N}(\bm{k})$. In fact, as shown in Appx.~\ref{appx:asym}, the imaginary part of the off-diagonal SW tensor element $\mathcal{A}(\bm{k})\equiv \Im N^{(21)}$ has to be an odd function of the wave vector $\bm{k}$ which means, in general, $\mathcal{A}(-\bm{k}) = -\mathcal{A}(\bm{k})$. In other words, depending on the system at hand, this term has to be either zero or lead to a dispersion asymmetry. Because of this, it denoted as the (here unitless) \textit{magnetochiral} stiffness field, which resonantly couples the two independent components (index 1 and 2) of the dynamical magnetization. Furthermore, one can show  that this field is the only possible one leading to an asymmetry, as all other terms have to be even in $\bm{k}$ (see again Appx.~\ref{appx:asym}). Therefore, the dispersion asymmetry for plane waves can be given as 
\begin{equation}
    \Delta\omega(\bm{k}) = \omega(\bm{k}) - \omega(-\bm{k}) = 2\omega_M\mathcal{A}(\bm{k}).
\end{equation}
It is clear, that the real challenge of obtaining the spin-wave dispersion lies in obtaining exactly the effective SW tensor elements $N_\nu^{(ij)}$. However, we will show here that, indeed it is possible to obtain each of these terms individually in a fully numerical approach.

\section{Method}\label{sec:methods}

After having introduced the general ingredients of a dynamic-matrix approach and having discussed the general form of the spin-wave dispersion, we want to address the issue of hybridization. For this purpose, we outline how to numerically obtain the unhybridized spin-wave dispersion in terms of modes with a given spatial profile or spatial symmetry, solely based on the numerical dynamic-matrix approach. We will see that the individual terms of the general dispersion Eq.~\eqref{eq:general-dispersion} are a natural outcome of the presented methodology. As announced, we present this methodology for a propagating-wave dynamic-matrix approach to calculate the dispersion of propagating modes in certain waveguides, where the dynamic matrix $\vu{D}_k$ is only solved in a single cross section of the waveguide. However, the method can of course be applied to any dynamic-matrix approach. Again, for the readers convenience, expressions for the plane-wave magnetic tensors are found in the Appx.~\ref{appx:mattensor}.

Modes with a given spatial profile or symmetry (\textit{e.g.} "the first perpendicular-standing mode" or "the first radial mode") are in some works simply referred to as the actual \textit{spin-wave modes}. Recall, that numerically diagonalizing the dynamic matrix intrinsically outputs the (already hybridized) normal modes of the given systems which may be mixtures of these spin-wave modes. In most theoretical works, these are taken synonymous with the completely unpinned (or partially pinned) exchange modes of the system which are the normal modes when no dynamic dipolar fields are present. Since, in general, non-dipolar interactions may also include anisotropy, DMI, and so forth, and, in order to avoid confusion, here and henceforth, we will denote these modes as \textit{non-dipolar} modes. To proceed further, the dynamic matrix is projected onto a basis of these modes. The following strategy has already been widely used in many analytical theories of spin-wave propagation for systems where analytical expressions for the spatial mode profiles are known (see for example Refs.~\citenum{kalinikosTheoryDipoleexchangeSpin1986}, \citenum{tacchiStronglyHybridizedDipoleexchange2019}, \citenum{gladiiFrequencyNonreciprocitySurface2016a} or \citenum{grassiSlowWaveBasedNanomagnonicDiode2020}). Let us note that it gives approximately correct results only when the spatial dependence of both the mode magnitude and precession ellipticity is not changed considerably by the influence of the dynamic dipolar fields. We shall later see an example where this method fails.

In a general case, finding the non-dipolar spectrum analytically is cumbersome, and may involve spatial mode profiles which do not have any closed analytical form (yet). For this purpose, we shall obtain the non-dipolar spectrum numerically as solutions of the eigenvalue problem
\begin{equation}\label{eq:nondipolarproblem}
    \omega^{(\mathrm{ND})}_\nu (k) \, \bm{s}_{\nu k} = \omega_M \vu{D}_{k}^{(\mathrm{ND})} \bm{s}_{\nu k}
\end{equation}
with $\vu{D}_{k}^{(\mathrm{ND})}$ being the dynamic matrix of given magnetic specimen with known equilibrium, including all interactions, but without the dynamic dipolar fields represented by $\vu{N}^{(\mathrm{dip})}_k$. Here, $\bm{s}_{\nu k}$ are the complex-valued spatial profiles of the non-dipolar modes and $\nu$ now denotes some lateral mode index which counts the different dispersion branches. Note that the dipolar equilibrium field should still be included in the equilibrium field $h_0$. Specifically, this matrix may also contain interactions such as anisotropy or DMI. The mode pinning of the basis modes is determined by the boundary condition Eq.~\eqref{eq:boundarycond} and possible surface anisotropy. Recall that the eigenvectors of the spatially discretized Eq.~\eqref{eq:nondipolarproblem} have $2n$ components. We can construct a basis of non-dipolar modes using the orthonormal basis vector fields
\begin{equation}
    \bm{g}_{\nu k}^{(1)} = \frac{1}{\sqrt{\mathcal{N}^{(1)}_{\nu k}}}\mqty(\bm{s}_{\nu k}\cdot \bm{e}_1 \\ 0 ), \quad \bm{g}_{\nu k}^{(2)} = \frac{1}{\sqrt{\mathcal{N}^{(2)}_{\nu k}}}\mqty(0 \\  \bm{s}_{\nu k}\cdot \bm{e}_2 )
\end{equation}
with $\bm{e}_{1,2}(\bm{r})$ being the basis vectors locally orthogonal to the equilibrium magnetization $\bm{m}_0(\bm{r})$. In this sense, $\bm{g}_{\nu k}^{(1,2)}$ allows us to treat the two components of the mode profiles individually. This is important, because at a later point when the dipolar interaction is included, it may change the precession ellipticity, \textit{i.e.} the ratio between the dynamical components. The normalization factors $\mathcal{N}_{\nu k}^{(1,2)}$ are chosen with respect to the cross-section area (or full volume) in the sense
\begin{equation}
    \mathcal{N}_{\nu k}^{(i)} = \int_S \mathrm{d}S^\prime \,\abs{ \bm{s}_{\nu k}(\bm{\rho}^\prime) \cdot \bm{e}_i (\bm{\rho}^\prime)}^2 \quad \text{with} \quad S=A,V.
\end{equation}
The normalization with respect to either cross-section area $A$ or magnetic volume $V$ has to be done depending on whether one uses a propagating-wave- or a full three-dimensional dynamic-matrix approach, \textit{i.e.} depending on the number of dimensions in which the magnetic medium is discretized.
Now, we are ready to calculate the matrix elements of the full dynamic matrix (including dipolar interaction) in the basis spanned by $\bm{g}_{\nu k}^{(1,2)}$ for all $\nu$ and have 
\begin{widetext}
\begin{equation}\label{eq:dynmatblocks}
   \vu{D}_k \ \hat{=}\ \mqty(\dmat{\vu{C}_{\nu_1, k},\vu{C}_{\nu_2, k},\ddots,\vu{C}_{n, k}})  + \mqty(\admat{ \strut,\vu{W}_{\nu\nu^\prime, k},\strut,\vu{W}_{\nu^\prime\nu, k},\strut})
\end{equation}
\end{widetext}
with the diagonal blocks
\begin{equation}
    \vu{C}_{\nu, k} = \mqty(C_{\nu, k}^{(11)} & C_{\nu, k}^{(12)} \\ C_{\nu, k}^{(21)} & C_{\nu, k}^{(22)}) 
\end{equation}
with
\begin{equation}\label{eq:diagonalblocks}
    C_{\nu, k}^{(ij)} = \int\mathrm{d}A' \, \left(\bm{g}_{\nu k}^{(i)}\right)^*\boldsymbol{\cdot}  \vu{D}_k \boldsymbol{\cdot} \bm{g}_{\nu k}^{(j)}
\end{equation}
and the off-diagonal blocks $\vu{W}_{\nu\nu^\prime, k}$ accordingly. As the full dynamic matrix has to be Hermitian it is clear that $\vu{W}_{\nu\nu^\prime, k}=\vu{W}_{\nu^\prime\nu, k}^\dagger$. These off-diagonal blocks are responsible for the mixing/hybridization of different spin-wave (non-dipolar) modes. The new unhybridized (or unperturbed) dipole-exchange spectrum in terms of modes with the old mode indices $\nu$ can now be obtained by setting all off-diagonal blocks to zero. In this zeroth order approximation, the dispersion is simply given by the eigenvalues of the diagonal blocks
\begin{equation}
    \det(\vu{D}_k^{(0)}-\omega^{(0)}\vu{I}_{2n}) = \prod\limits_{\nu}\det(\vu{C}_{\nu k}-\omega^{(0)}\vu{I}_{2})
\end{equation}
which finally yields the explicit unperturbed dipole-exchange dispersion neglecting hybridization
\begin{equation}\label{eq:zerothorderdisp}
\begin{split}
        \omega_\nu^{(0)}(k) = & \frac{1}{2}\Bigg(C_{\nu, k}^{(11)} + C_{\nu, k}^{(22)}\Bigg) + \\ & \sqrt{C_{\nu, k}^{(12)}C_{\nu, k}^{(21)} -  \mqty[\frac{i}{2}\mqty(C_{\nu, k}^{(11)} - C_{\nu, k}^{(22)})]^2 }
\end{split}
\end{equation}
Accordingly, the first order perturbation (considering two hybridizing branches $\nu_1$ and $\nu_2$) can be obtained by 
\begin{equation}
\det\mqty(\vu{C}_{\nu_1 k} -\omega^{(1)}\vu{I}_{2} & \vu{W}_{\nu_1\nu_2 k} \\ \vu{W}_{\nu_1\nu_2 k}^\dagger & \vu{C}_{\nu_2 k}-\omega^{(1)}\vu{I}_{2}) = 0.
\end{equation}
However, of course the exact dipole-exchange spectrum obtained by strictly diagonalizing the full dynamic matrix $\vu{D}$ numerically. 
Considering the diagonal blocks of Eq.~\eqref{eq:diagonalblocks} it is clear that 
the zeroth-order dispersion in Eq.~\eqref{eq:zerothorderdisp} is formally identical to the general form of the spin-wave dispersion in Eqs.~\eqref{eq:general-dispersion-intro} and \eqref{eq:general-dispersion} by identifying
\begin{subequations}\label{eq:tensor-components}
\begin{align}
   \mathcal{A}_{\nu}(k)&=  \Im N^{(21)}_{\nu k}  = \frac{1}{2}\Bigg(C_{\nu, k}^{(11)}+ C_{\nu,k}^{(22)}\Bigg)\\
    \mathcal{B}_{\nu}(k)&= N^{(11)}_{\nu k} + h_0  = -iC_{\nu,k}^{(21)}\\
    \mathcal{C}_{\nu}(k)&= N^{(22)}_{\nu k} + h_0  = iC_{\nu,k}^{(12)}\\
    \mathcal{D}_{\nu}(k)&= \Re N^{(21)}_{\nu k}  = \frac{i}{2}\mqty(C_{\nu, k}^{(11)} - C_{\nu, k}^{(22)}).
\end{align}
\end{subequations}
Let us stress the point that the matrix elements $ C_{\nu, k}^{(ij)}$ are obtained directly from numerics. As a result, we are now able to fully numerically calculate the unperturbed dipole-exchange spectrum in an arbitrary magnetic specimen and can even disentangle the individual terms of the dispersion. Note that, the above-described procedure can of course also be done in terms of the already hybridized mode profiles, \textit{i.e.} in terms of normal modes, by using the eigenvectors $\bm{m}_{\nu^\prime k}$ of the full matrix $\vu{D}_k$ as a basis, instead of the non-dipolar profiles $\bm{s}_{\nu k}$. Obviously, taking this route changes the indexing of the modes and renders the dynamic matrix in Eq.~\eqref{eq:dynmatblocks} block diagonal from the beginning, \textit{i.e.} all mixing blocks $\vu{W}_{\nu\nu^\prime}$ vanish.

\section{Applications}\label{sec:applications}

In the following, we would like to showcase the presented methodology with three different examples. First, we will show how to disentangle hybridized modes in a transversally magnetized rectangular waveguide. This examples also highlights one of the short comings of this approach, namely the assumption that the spatial dependencies of both the mode magnitude and of the ellipticity are not affected by dynamic dipolar fields. After this, we will present how the different dynamic stiffness fields can be calculate separately, and, furthermore, how to resolve contributions from different magnetic interactions.

\subsection{Unhybridized dispersion transversally magnetized waveguide}

One of the standard magnetic elements considered in various spin-wave transport experiments is the transversally-magnetized rectangular waveguide.\cite{sebastianNonlinearEmissionSpinWave2013,bracherTimePowerdependentOperation2014,hulaNonlinearLossesMagnon2020,schultheissTimeRefractionSpin2021} Here, we consider a soft-magnetic waveguide of \SI{800}{\nano\meter} width and \SI{50}{\nano\meter} thickness shown in Fig.~\figref{fig:rect}{a}, with typical material parameters of Ni$_{80}$Fe$_{20}$ (permalloy) summarized in Tab.~\ref{tab:matparam}.
\begin{table}[h!]
\caption{\label{tab:matparam}Parameters used for micromagnetic modeling.}
\begin{ruledtabular}
\begin{tabular}{ll}
		exchange stiffness, $A_\mathrm{ex}$ & \SI{13}{\pico\joule/\meter}\\
		saturation, $M_\mathrm{s}$ & \SI{796}{\kilo\ampere/m}\\
		reduced gyromagnetic ratio, $\gamma/2\pi$ & \SI{28}{\giga\hertz/\tesla}\\
		bulk DMI constant (only Sec.~\ref{sec:dmi}), $D$ & \SI{1}{\milli\joule/\square\meter}
\end{tabular}
\end{ruledtabular}
\end{table}

The waveguide is magnetized in an S state which is stabilized by a static transversal field of $\mu_0 H_\mathrm{ext} = \SI{50}{\milli\tesla}$. Such an equilibrium state consists of a large transversally magnetized domain in the center of the waveguide (here: bulk domain) and two small edge domains on either side where the magnetization tries to align parallel with the waveguide edges [see Figs.~\figref{fig:rect}{a,b}]. The spin waves propagating along such a waveguide can be divided into the edge-domain modes, which propagate only in the edge domains, the bulk mode (or fundamental, or quasi-uniform) mode which exhibits a mostly homogeneous mode profile within the waveguide cross section, as well as higher-order transversal modes which exhibit a standing-wave character along the width of the waveguide. In Fig.~\figref{fig:rect}{c}, we show the spatial profiles of these \textit{non-dipolar} modes, calculated, as described in the previous section, according to Eq.~\eqref{eq:nondipolarproblem}.

It is well-known that the modes with the aforementioned spatial profiles are actually not normal modes of this particular system. Instead, for example the bulk mode and the higher-order transversal modes are hybridized by dynamic dipolar fields. In Fig.~\figref{fig:rect}{d}, we show the actual dispersion for the different hybridized normal modes as dashed lines, calculated by numerically diagonalizing the full dynamic matrix $\vu{D}_k$ of the system. On top of it, as solid lines, we show the unhybrized dispersion in terms of the non-dipolar modes (edge-domain, bulk, transversal) calculated using the numerical scheme presented in this paper. The bulk mode is crossing several higher-order transversal modes and, as can be seen in Fig.~\figref{fig:rect}{e}, hybridizes with some of them.\footnote{Whether two modes can hybridize depends on the similarity of their spatial symmetries. For example, the bulk mode only hybridizes with modes with an even number of nodal lines across the width of the waveguide.} We see that, in fact, one can completely recover the unhybridized dispersion in terms of non-dipolar modes from numerics. As a short coming, there are significant deviations in the frequencies of the edge-domain modes, when comparing the non-dipolar with the normal modes. This is due to the fact that the dynamic dipolar fields influence the localization as well as the spatial dependence of the ellipticity of these modes. Note, however, that comparing numerical results for edge-domain modes with experimental data can be ambiguous, as they are strongly influenced by the quality of the edge.

\begin{figure}[h!]
    \centering
    \includegraphics{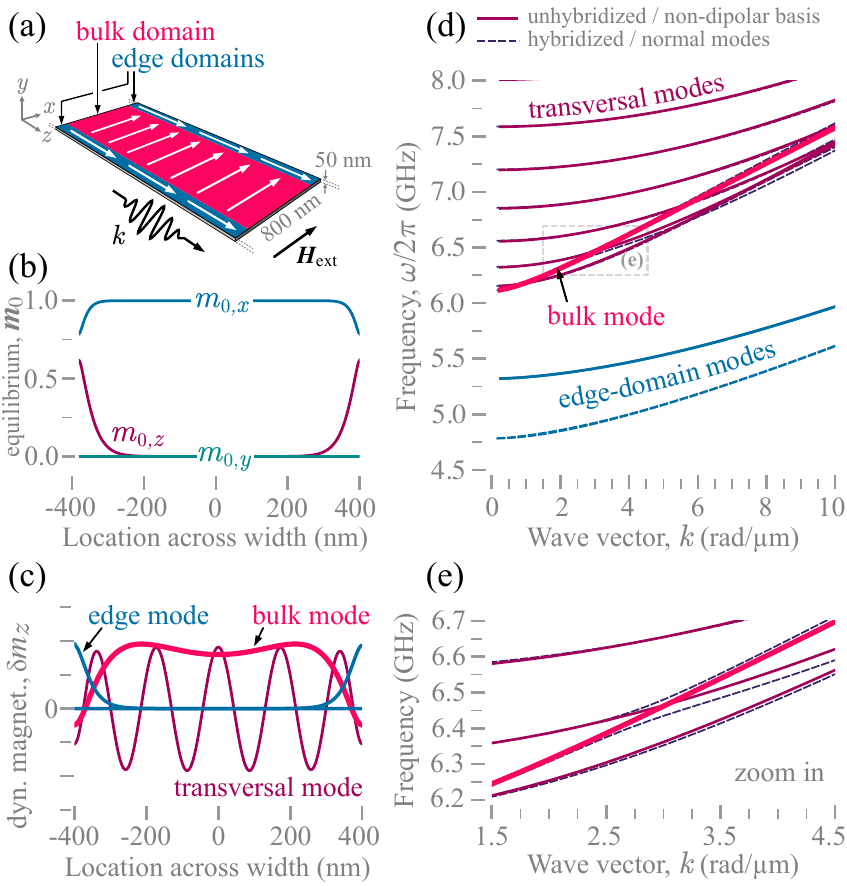}
    \caption{(a) Schematics of a soft-magnetic rectangular waveguide in the S state, with a transversally applied static external field of $\mu_0 H_\mathrm{ext} = \SI{50}{\milli\tesla}$. (b) Scan of the normalized equilibrium magnetization across the width of the waveguide. The equilibrium state has been obtained using an energy minimizer. (c) Scans of the lateral spatial profiles (only out-of-plane component) for different non-dipolar modes in the waveguide (see text). (d) Hybridized dipole-exchange spectrum (dashed) of the normal modes in the waveguides, overlayed with the unhybridized spectrum (solid) in terms of non-dipolar modes. (e) Zoom in of the same dispersion showing an anti-crossing of two different modes.} 
    \label{fig:rect}
\end{figure}

Albeit there is not strict analytical model for the spin waves in an S-state waveguide, both the bulk mode and the transversal modes can be approximately described using the dispersion of thin films provided by Kalinikos and Slavin in Ref.~\citenum{kalinikosTheoryDipoleexchangeSpin1986} by numerically determining the magnetic ground state and then extracting
\textit{approximate} values for the effective magnetic field and the effective width of the waveguide.\cite{hulaNonlinearLossesMagnon2020,schultheissTimeRefractionSpin2021}

\subsection{Identification of individual stiffness fields in magnetic nanotubes}

As a next example, we show that the above-presented methodology easily allows to individually determine the different components of the effective spin-wave tensor $N_{\nu k}^{(ij)}$ via Eq.~\eqref{eq:tensor-components}, which are connected to the dynamic stiffness fields $\mathcal{A}_\nu(k)$, $\mathcal{B}_\nu(k)$, $\mathcal{C}_\nu(k)$ and $\mathcal{D}_\nu(k)$ that make up the full spin-wave dispersion. As seen in Fig.~\figref{fig:tube}{a}, we consider a magnetic nanotube in the vortex state, with \SI{20}{\nano\meter} inner and \SI{30}{\nano\meter} outer radius and the same material parameters as in the previous section. The vortex state is stabilized using an azimuthal field $B_\phi=\SI{80}{\milli\tesla}$. The spin waves propagating along such a tube have already been discussed extensively in the literature by Otálora \textit{et al.},\cite{otaloraCurvatureInducedAsymmetricSpinWave2016,otaloraAsymmetricSpinwaveDispersion2017,otaloraFrequencyLinewidthDecay2018} and are known to exhibit a curvature-induced dispersion asymmetry, \textit{i.e.} $\mathcal{A}_\nu(k)\neq 0$, which originates from the dipolar interaction. Here, for the case of a thin-shell nanotube in the vortex state, $\mathcal{D}_\nu(k)\equiv 0$.\cite{otaloraCurvatureInducedAsymmetricSpinWave2016,kornerTwomagnonScatteringPermalloy2013} As a result, the dispersion reduces to 
\begin{equation}
    \omega_\nu (k) = \omega_M\left[\mathcal{A}_\nu(k) + \sqrt{\mathcal{B}_\nu(k)\mathcal{C}_\nu(k)}\right].
\end{equation}
The spatial profile of the modes in the tube is well-described by $\propto\exp[i(\nu\phi + kz)]$ where, $\nu\in\mathbb{Z}$ denotes the number of azimuthal periods. In Fig.~\figref{fig:tube}{b} we show the numerically obtained asymmetric dispersion of these modes together with the predictions provided by the theory of Ref.~\citenum{otaloraCurvatureInducedAsymmetricSpinWave2016}. Finally, the individual (unitless) stiffness fields obtained numerically and compared with the analytical model, are shown in Figs.~\figref{fig:tube}{c-e}. 

\begin{figure}[h!]
    \centering
    \includegraphics{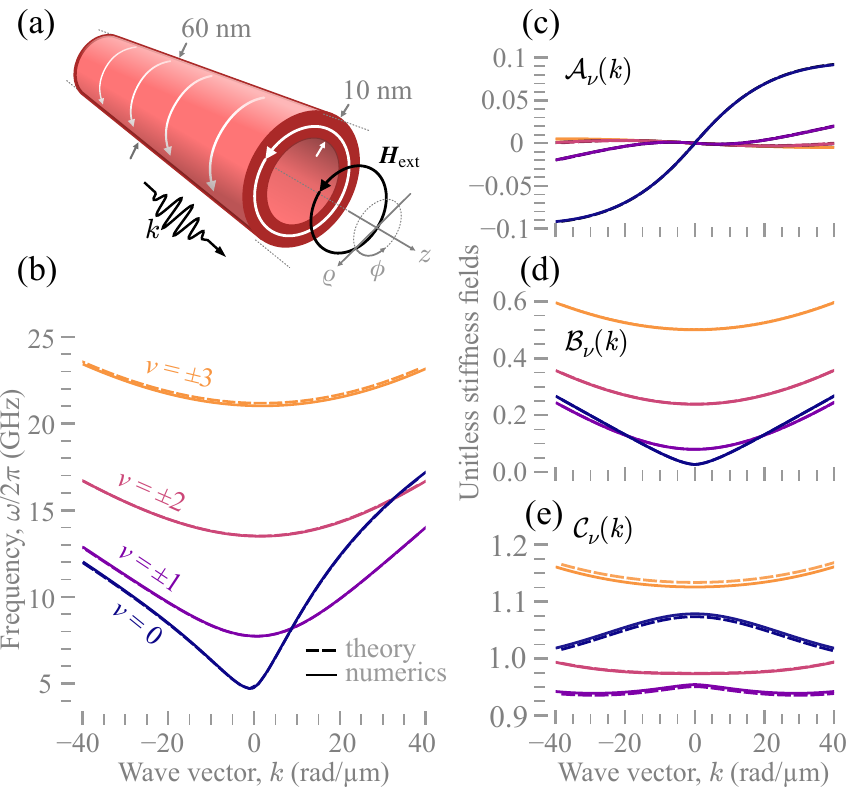}
    \caption{(a) Schematics of a thin-shell permalloy nanotube in the vortex state. (b) Asymmetric spin-wave dispersion of the different azimuthal modes [$\propto\exp(i\nu\phi)$] propagating along the tube. (c-e) Numerically obtained individual stiffness fields for the different modes, overlayed with the theoretical prediction according to Ref.~\citenum{otaloraCurvatureInducedAsymmetricSpinWave2016}.}
    \label{fig:tube}
\end{figure}

As can bee seen, the fully numerical methodology almost perfectly reproduces the theoretical predictions. The deviations in the stiffness field $\mathcal{C}_\nu(k)$ are insignificant as this (unitless) field is close to unity, \textit{i.e.} the overall influence on the full dispersion remains negligible. In agreement with our previous discussion on the general form of the spin-wave dispersion in Sec.~\ref{sec:genera-dispersion}, only the magnetochiral field $\mathcal{A}_\nu(k)$ is asymmetric (more precisely: odd) in $k$, and therefore, solely induces the dispersion asymmetry. In this case, this magnetochiral field couples the $z$ and the $\rho$ components of the dynamical magnetization $\delta\bm{m}(\bm{r},t)$, where as the $\mathcal{B}_\nu$ ($\mathcal{C}_\nu$) field is produced by the $z$ ($\rho$) component and exclusively acts on this specific component.

\subsection{Disentanglement of various contributions to the  dispersion asymmetry}\label{sec:dmi}
\begin{figure}[h!]
    \centering
    \includegraphics[width=8.5cm]{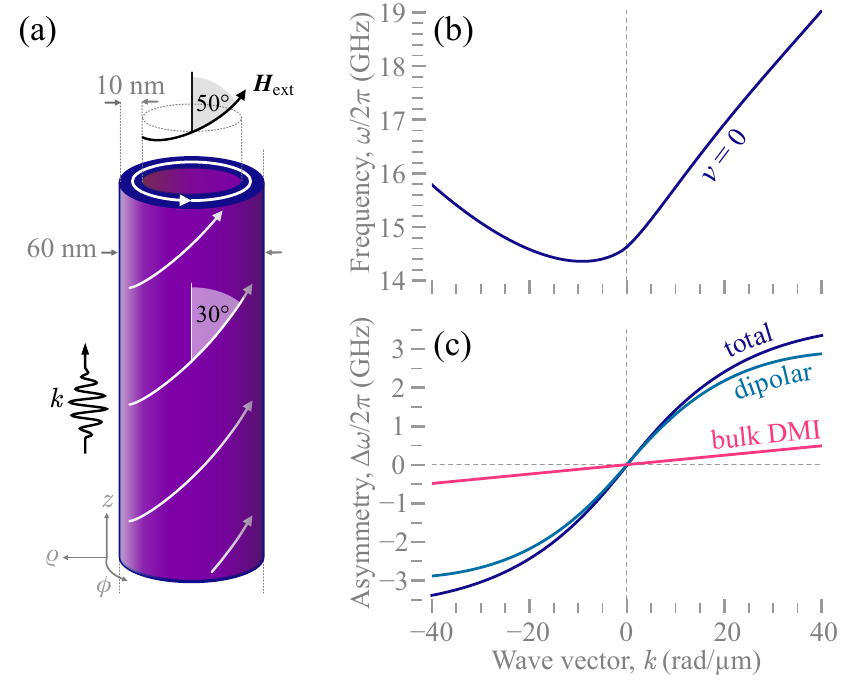}
    \caption{(a) Schematics of a thin-mantle magnetic nanotube in the helical state with bulk DMI. (b) Asymmetric dispersion of the lowest-order mode which has a homogeneous lateral mode profile. (c) Total dispersion asymmetry and contributions of the individual magnetic interactions as a function of wave vector.}
    \label{fig:dmi}
\end{figure}
Since the magnetic tensor $\vu{N}$, and therefore the full dynamic matrix $\vu{D}$, are additive in the various magnetic interactions, of course, using this approach, the individual elements $N_{\nu k}^{(ij)}$ of the effective spin-wave tensors, and, with these, the different stiffness fields, can be calculated for each interaction individually, without any difficult coding. Therefore, as a last application of this approach, we show that it is easily possible to disentangle different contributions to any dispersion asymmetry. Now we consider a magnetic nanotube, magnetized in the helical state [see Fig.~\figref{fig:dmi}{a}], with the same material parameters as before, however additionally with a non-zero bulk DMI constant of $D=\SI{1}{\milli\joule/\square\meter}$ (for expressions of the magnetic tensor including bulk DMI, see Appx.~\ref{appx:mattensor}). The helical state is stabilized using a field in both the $z$ and the $\phi$ direction, according to
\begin{equation}
    \bm{H}_\mathrm{ext} =  H_\mathrm{ext}\left[\sin(\Theta_H)\bm{e}_\phi + \cos(\Theta_H)\bm{e}_z\right]
\end{equation}
with the angle of the field with the $z$ axis $\Theta_H = 50^\circ$ and the magnitude $\mu_0 H_\mathrm{ext}=\SI{200}{\milli\tesla}$. The equilibrium magnetization in the helical state has the same mathematical form as the external field but encloses an angle of $\Theta_M = 30^\circ$ with the $z$ axis.

In this case, it is strongly expected that both the curvature-induced magnetic charges,\cite{salazar-cardonaNonreciprocitySpinWaves2021} as well as the DMI contribute to a dispersion asymmetry.\cite{cortes-ortunoInfluenceDzyaloshinskiiMoriya2013} This is a particularly complicated example for which there is no analytical theory available yet. In Fig.~\figref{fig:dmi}{b}, we show the dispersion for the lowest order ($\nu=0$) mode, which exhibits a homogeneous profile within the nanotube cross section. As can been seen, this dispersion branch exhibits a strong dispersion asymmetry. Finally, by obtaining the magnetochiral field $\mathcal{A}_0(k)$ for each interaction individually, one can disentangle their contributions to the dispersion asymmetry $\Delta\omega_0(k) = 2\omega_M\mathcal{A}_0(k)$ in Fig.~\figref{fig:dmi}{c}. As expected, the bulk DMI contributes linearly as in thin films, whereas the dipolar interaction contributes nonlinearly. Note, that both contributions are odd functions of the wave vector $k$. 

\section{Conclusion}

Based on a dynamic-matrix approach, we presented a methodology to reverse engineer spin-wave dispersions in a fully numerical fashion. After having discussed the general form of any spin-wave dispersion and their individual ingredients, namely the stiffness fields, we have shown that these fields can be calculated without having knowledge about the mode profiles beforehand. We have shown that each stiffness field in a dispersion can be obtained individually, for each magnetic interaction separately. In particular, it becomes possible to disentangle different contributions which lead to an asymmetric dispersion, by evaluating the magnetochiral stiffness field, which is shown to be the only one that can lead to a dispersion asymmetry, for any interaction.

Furthermore, we have shown that the presented methodology allows to easily disentangle dipole-dipole hybridized modes, something which, in most cases, required an analytical or semi-analytical theory before. This was done by evaluating the unperturbed/unhybridized dispersion in terms of numerically-obtained non-dipolar modes, which works well for modes whose spatial profile is not changed significantly by the dynamic dipolar fields.

Although, we have presented explicit examples using a finite-element propagating-wave dynamic-matrix approach, the method can be applied for any dynamic-matrix approach. This means that the presented methodology is suitable for a large number of different geometries and equilibrium states. We believe, that this work is of particular interest for the study of spin waves in complex mesoscopic systems where analytical models are not available yet. Moreover, it might help in the development and verification of new analytical theories. 

\section*{Acknowledgements}
We are very thankful to Pedro Landeros for fruitful discussions. Financial support by the Deutsche Forschungsgemeinschaft within the programs KA 5069/1-1 and KA 5069/3-1 is gratefully acknowledged.


\section*{Appendix}

\begin{appendix}

\section{Contributions to the magnetic tensor}\label{appx:mattensor}

Here, we briefly showcase the magnetic tensors for the magnetic interactions considered in this manuscript.
The magnetic tensors $\vu{N}$ for dipolar and exchange interaction are defined by  
\begin{align}
    \hat{\mathbf{N}}^{(\mathrm{dip})} \bm{m}_\nu & = \nabla \phi_\nu\\
    \hat{\mathbf{N}}^{(\mathrm{ex})} &= -\lambda_\mathrm{ex}^2\nabla^2. 
\end{align}
with $\phi_\nu$ being the dipolar potential generated by the mode $\nu$ and $\lambda_\mathrm{ex}$ is the exchange length of the material.\cite{korberFiniteelementDynamicmatrixApproach2021} In the present paper, we used a propagating-wave dynamic-matrix approach for infinitely extended waveguides where the equilibrium magnetization is assumed to be translationally invariant along the $z$ direction. In this case, the magnetic tensors are transformed to plane-wave tensors $\vu{N}_k$ by 
\begin{equation}\label{eq:trafo}
    \vu{N} \rightarrow \vu{N}_k = e^{-ikz}\vu{N}e^{ikz}
\end{equation}
as described for example in Ref.~\citenum{henryPropagatingSpinwaveNormal2016} or in our previous work Ref.~\citenum{korberFiniteelementDynamicmatrixApproach2021}. Here, we additionaly present the plane-wave tensor for the bulk Dzyaloshinskii-Moriya interaction. One starts with the three-dimensional magnetic tensor for bulk DMI (in crystals with T symmetry),\cite{cortes-ortunoInfluenceDzyaloshinskiiMoriya2013,bogdanovNewLocalizedSolutions1995a} defined by
\begin{equation}
    \hat{\mathbf{N}}\bm{m}_\nu = -\frac{2D}{M_\mathrm{s}^2} \nabla \cross \bm{m}_\nu
\end{equation}
with the bulk DMI constant $D$.
By assuming the mode profile to be 
\begin{equation}
    \bm{m}(\bm{r})=\bm{\eta}_{\nu k} (x,y)e^{ikz}
\end{equation}
with some lateral profile $\bm{\eta}_{\nu k}$, and by performing the transformation Eq.~\eqref{eq:trafo} one arrives at 
\begin{equation}\label{eq:dmi-operator}
\begin{split}
        \Hat{\mathbf{N}}_k \bm{\eta}_{\nu k} &= -\frac{2D}{M_\mathrm{s}^2} \left(k\vu{\Sigma}\bm{\eta}_{\nu k} + \nabla\times\bm{\eta}_{\nu k}  \right) \\
        & = -\frac{2D}{M_\mathrm{s}^2} \left(k\vu{\Sigma} + \nabla\times\hdots  \right)\bm{\eta}_{\nu k}
\end{split}
\end{equation}
which now only acts on the lateral profiles $\bm{\eta}_{\nu k}(x,y)$ of the plane waves with a given wave number $k$. The operator $\vu{\Sigma}$ is given in the lab system as
\begin{equation}
    {\hat{\mathbf{\Sigma}}} = \mqty( 0 & -i & 0 \\ i & 0 & 0 \\ 0 & 0 & 0 ).
\end{equation}
The reader can convince themself that the plane-wave bulk DMI tensor defined by Eq.~\eqref{eq:dmi-operator} is still self adjoint.

\section{Contributions to the dynamic matrix}\label{appx:dynmat}

The dynamic matrix, which is numerically diagonalized to obtain the spin-wave spatial profiles and frequencies, is expressed as 
\begin{equation}
    \hat{\mathbf{D}} = i\hat{\bm{\Lambda}}  \hat{\mathbf{R}} \vu{\Omega} \hat{\mathbf{R}}^\dagger
\end{equation}
with the rotation operator 
\begin{equation}
    \hat{\mathbf{R}} = \begin{pmatrix}
    \bm{e}_1 \cdot \bm{e}_x & \bm{e}_1 \cdot \bm{e}_y & \bm{e}_1 \cdot \bm{e}_z \\ \bm{e}_2 \cdot \bm{e}_x & \bm{e}_2 \cdot \bm{e}_y & \bm{e}_2 \cdot \bm{e}_z 
    \end{pmatrix},
\end{equation}
which transforms vectors from the lab system $\{\bm{e}_x, \bm{e}_y, \bm{e}_z\}$ to the basis $\{\bm{e}_1, \bm{e}_2\}$ locally orthogonal to the direction of the equilibrium magnetization $\bm{m}_0(\bm{r})$. This basis can be obtained locally by 
\begin{equation}
    \bm{e}_1 = -  \frac{\bm{m}_0\times\bm{e}_2}{\vert \bm{m}_0\times\bm{e}_2 \vert },\qquad \bm{e}_2 = \frac{\bm{e}_z\times\bm{m}_0}{\vert \bm{e}_z\times\bm{m}_0 \vert }.
\end{equation}
Note, that the operator $\vu{R}$ is unitary for vectors orthogonal to $\bm{m}_0$. This holds true in particular for all valid mode profiles, since $\delta\bm{m}\perp\bm{m}_0$. The operator $\vu{\Lambda}$ represents the cross-product operator $\bm{m}_0 \times \hdots$ and, in the local basis, takes the simple form
\begin{equation}
    \hat{\bm{\Lambda}}= \begin{pmatrix} 0 & -1  \\ 1 & 0\end{pmatrix}.
\end{equation}
The operator $\vu{\Omega}$ is defined as in the main text.

\section{Asymmetry of effective spin-wave tensor elements}\label{appx:asym}
In this section we briefly discuss the asymmetry of the matrix elements of the effective spin-wave tensors $\vu{N}_\nu$ upon reversal of the wave vector. We will see the dispersion asymmetry can only come from the imaginary part $\Im N^{(12)}$ of the effectice spin-wave tensor written in the local basis $\{\bm{e}_1, \bm{e}_2\}$, independent of the magnetic interaction at hand.

Let us assume that the mode profiles are plane waves and the effective spin-wave tensors $\vu{N}_\nu \equiv \vu{N}(\bm{k})$ are defined by 
\begin{equation}\label{eq:def-plane-wave-effective-tensors}
  -  \delta\bm{h} = \vu{N}\cdot \delta\bm{m} = \sum\limits_{\bm{k}} \vu{N}(\bm{k})\bm{a}_{\bm{k}}(t) e^{i\bm{k}\bm{r}}
\end{equation}
with $\delta\bm{h}$ again being the unitless effective field generated by the dynamic magnetization $\delta\bm{m}$ and $\bm{a}_{\bm{k}}$ being a time-dependent complex amplitude vector of the respective mode. In the case of pure plane waves, we have two important properties:
\begin{enumerate}
    \item[(i)] Since the full magnetic tensor is Hermitian, so have to be the effective spin-wave tensors, $\vu{N}^\dagger (\bm{k}) =(\vu{N}^*)^\mathrm{T} (\bm{k}) = \vu{N}(\bm{k})$. This holds true for all effective spin-wave tensors.
    \item[(ii)] The dynamic effective field $\delta\bm{h}$ has to be real-valued. Then, from Eq.~\eqref{eq:def-plane-wave-effective-tensors}, it follows that $\vu{N}^*(\bm{k}) = \vu{N}(-\bm{k})$.
\end{enumerate}
Let us know consider an effective spin-wave tensor written in the local basis
\begin{equation}
  \vu{N}(\bm{k})=\mqty(N_{11}(\bm{k}) & N_{12}(\bm{k})  \\ N_{21}(\bm{k})  & N_{22}(\bm{k}) ).
\end{equation}
For readibility, in this section, we will write the indices as subscript.
From (i), it follows directly that the diagonal components $N_{11}, N_{22}$ have to be real. Then, from (ii) it follows that they have to be even in $\bm{k}$, i.e.
\begin{align}
        N_{11}(\bm{k}) & \overset{(\text{i})}{=} N_{11}^*(\bm{k}) \overset{(\text{ii})}{=}  N_{11}(-\bm{k}) \\
                N_{22}(\bm{k}) & = N_{22}^*(\bm{k}) = N_{22}(-\bm{k}) 
\end{align}
Now, let us focus on the off-diagonal element $N_{21}(\bm{k})=N_{21}^*(\bm{k})$. Splitting it into real and imaginary part, we can write
\begin{equation}\label{eq:eq1}
\begin{split}
        N_{21}^*(\bm{k}) & = \Big\{\Re\big[N_{21}(\bm{k})\big] + i\Im\big[N_{21}(\bm{k})\big]\Big\}^*\\ & = \Re\big[N_{21}(\bm{k})\big] - i\Im\big[N_{21}(\bm{k})\big].
\end{split}
\end{equation}
Now, from (ii), we also have
\begin{equation}\label{eq:eq2}
\begin{split}
        N_{21}^*(\bm{k}) & = N_{21}(-\bm{k}) \\ &= \Re\big[N_{21}(-\bm{k})\big] + i\Im\big[N_{21}(-\bm{k})\big].
\end{split}
\end{equation}
Comparing Eqs.~\eqref{eq:eq1} and \eqref{eq:eq2} and using the fact that real and imaginary part and linearly indepedent, we can conclude
\begin{align}
        \Re\big[N_{21}(\bm{k})\big] & = \Re\big[N_{21}(-\bm{k})\big] \\
                \Im\big[N_{21}(\bm{k})\big] & = -\Im\big[N_{21}(-\bm{k})\big] 
\end{align}
which means that, in fact, the only component of the effective spin-wave tensor which can (and actually has to) produce a dispersion asymmetry, is the imaginary part of the off-diagonal elements $\Im N_{21} = -\Im N_{12} $. In the main text, this quantity is reffered to as the (unitless) magnetochiral field $\mathcal{A}$. 

This discussion can also be generalized for spin waves which are not 3D plane waves, for example the modes in a rectangular waveguide with small cross section might be characterized by the wave vector $k$ and two lateral mode indices $n$ and $\ell$. The same arguments as above can be made in case the mode profiles can be written as 
\begin{equation}
    \delta\bm{m} = \sum\limits_{\bm{\nu},\bm{q}} \big[{m}^{(1)}_{\bm{\nu},\bm{q}}(\bm{\rho})\,\bm{e}_1 + i{m}^{(2)}_{\bm{\nu},\bm{q}}(\bm{\rho})\,\bm{e}_2 \big] e^{-i\omega t}e^{i\bm{q}\bm{s}}.
\end{equation}
Here, ${m}^{(1)}_{\bm{\nu},\bm{q}}$ and ${m}^{(2)}_{\bm{\nu},\bm{q}}$ are real-valued functions and represent the \textit{standing-wave} part of the mode profiles which depends on the generalized coordinates $\bm{\rho}=\{\rho_1,...,\rho_d\}$ and can be characterized using the mode indices $\bm{\nu}=\{\nu_1,...,\nu_d\}$. Similarly, the exponential \textit{propagating} part depends on the generalized coordinates $\bm{s}=\{s_1,...,s_p\}$ and is characterized by the generalized wave vector $\bm{q}=\{q_1,...,q_p\}$. Obviously, the sum $d+p$ has be equal to the number of degrees of freedom. Then, one can show the $\Im N_{21}$ has to be an odd function of all components of $\bm{q}$, or not depend on a specific component at all.
\end{appendix}
%

\end{document}